\documentclass[10pt]{iopart}

\usepackage{iopams}  

\usepackage[colorlinks=true, citecolor=red, urlcolor=blue ]{hyperref}
\usepackage[normalem]{ulem}
\usepackage{xcolor}
\usepackage{esvect}
\usepackage{cancel}
\usepackage[T1]{fontenc}
\usepackage[export]{adjustbox}
\usepackage{graphicx}
\usepackage{placeins}

\usepackage{multirow}
\usepackage{hhline,booktabs}

\usepackage{bibunits}
\usepackage{setspace}
\usepackage{blindtext}
\usepackage{graphicx}
\usepackage{bm}
\usepackage{mathrsfs} 
\usepackage{hyperref}
\usepackage{geometry}
\usepackage{amsthm}

\usepackage{epstopdf} 
\usepackage{graphicx}
\usepackage{lipsum} 

\begin{document}

\title{Magnetic Field Detection Using a Two-Qubit System Under Noisy Heisenberg Interaction}
\author{George Biswas *} 
\ead{georgebsws@gmail.com}

\address{Department of Physics, Tamkang University, Tamsui Dist., New Taipei 25137, Taiwan, ROC}
\address{Center for Advanced Quantum Computing, Tamkang University, Tamsui Dist., New Taipei 25137, Taiwan, ROC}

\author{Sayan Sengupta}

\address{Department of Physics, National Institute of Technology Sikkim, Ravangla, South Sikkim 737 139, India}

\author{Anindya Biswas *}
\ead{anindya@nitsikkim.ac.in}

\address{Department of Physics, National Institute of Technology Sikkim, Ravangla, South Sikkim 737 139, India}

\date{\today}

\begin{abstract}

We propose a method to design a magnetic field detector using a noisy two-qubit system. The system evolves under a noisy Heisenberg interaction Hamiltonian, and we investigate its behavior by calculating both the $l_1$-norm of quantum coherence and the return probability in the presence and absence of an external magnetic field. We allow for decoherence modeled by quasi-static charge noise in the exchange coupling of the two-qubit system and find that, while the magnetic field does not significantly influence the decoherence process, it introduces a distinct oscillation in the return probability over time. Importantly, the oscillation frequency is directly proportional to the strength of the applied magnetic field, providing a clear signature that can be used for magnetic field detection. These results point towards the feasibility of realizing a practical quantum-based magnetic field detector, with the ability to operate under noisy conditions while maintaining sensitivity to the field strength.

\end{abstract}

\vspace{2pc}
\noindent{\it Keywords}: Quantum Coherence, Return Probability, Magnetic field sensing, Heisenberg Interaction, Quasi-Static Charge Noise.

\vspace{2pc}
\footnotesize
\noindent{* Authors to whom any correspondence should be addressed.}
\normalsize

\maketitle

\section{Introduction}
\label{intro} 

Quantum magnetic field detectors/sensors~\cite{Bal2012,PhysRevLett.116.240801,Esat2024} and other quantum sensors~\cite{RevModPhys.89.035002} are advanced devices that use the principles of quantum mechanics to measure physical quantities with incredible precision. Unlike traditional (classical) sensors, quantum sensors can detect even the tiniest signals. However, quantum sensors have their challenges. One major issue is noise~\cite{mi14101823,GUO2024115323,https://doi.org/10.1002/qute.202300391}, which refers to unwanted disturbances that can interfere with the sensor's ability to measure accurately. Quantum sensors are highly sensitive, so they can easily pick up noise from the environment, which reduces their effectiveness.
For a quantum sensor to work well, it needs to respond strongly to the signals we want to measure, while at the same time being less affected by noise. This is a tricky balance to achieve~\cite{
doi:10.1126/science.aaw4278,PhysRevLett.127.080504,doi:10.2514/1.J062707}. Because of this, studying how noise affects quantum sensors is essential. Understanding and controlling this noise is key to improving the performance of quantum sensors and making them practical for real-world use. In this paper, we explore the feasibility of using a two-body spin system~\cite{PhysRevResearch.2.013061,wang2024neuralnetworkapproachtwobody} as a device for magnetic field detection.

\noindent Our approach involves measuring the $l_1$-norm of quantum coherence~\cite{RevModPhys.89.041003,PhysRevLett.115.020403,PhysRevA.100.012315,PhysRevA.97.022342,bu2016average,PhysRevA.93.062105,PhysRevA.93.032125,Biswas_2023} and return probability~\cite{Krapivsky_2018,e24050584} of the two-spin system after it has evolved for a certain period in the presence of a magnetic field. Mathematically $l_1$-norm of quantum coherence (denoted as $C_{l_1}$) is the sum of absolute values of the off diagonal elements of the time evolved density matrix $\rho(t)$,
\begin{equation}
    C_{l_1}=\sum_{i\ne j}|\rho_{ij}(t)|.
\end{equation}
{The $l_1$-norm is defined such that, for a single qubit state, it ranges between 0 and 1. However, for systems with more than one qubit, or for qutrits and qudits, the $l_1$-norm ranges between 0 and $n-1$, where $n$ is the dimension of the Hilbert space. In our case, we consider a two-qubit system, for which the Hilbert space has a dimension of $n=4$. Consequently, the $l_1$-norm of the two-qubit state ranges between 0 and $4-1$. For convenience, we prefer to work with a parameter that ranges between 0 and 1. Therefore, we normalize the $l_1$-norm by a factor of $n-1$.
\begin{equation}
    C_{l_1}=\frac{1}{n-1}\sum_{i\ne j}|\rho_{ij}(t)|.
\end{equation}}

\noindent Return probability (denoted as $P_{R}$) measures the fidelity or the overlap of the time evolved density matrix $\rho(t)$ with respect to the initial state $\psi(0)$, 
\begin{equation}
    P_{R}=\langle \psi(0)|\rho(t)|\psi(0)\rangle.
\end{equation}

\noindent During this evolution, the system is also subjected to environmental noise, which gradually reduces its coherence -- a process known as decoherence. By observing how the coherence ($C_{l_1}$) and return probability ($P_{R}$) behaves over time in the presence of a magnetic field, we can measure the magnetic field strength. In essence, the magnetic field influences how the noise affects the coherence and return probability dynamics of the spin system. By comparing dynamics with and without magnetic fields, we can develop a sensitive method for detecting magnetic fields, with the coherence and return probability dynamics of the spin system serving as key indicators.

\noindent Another aim of this study is to identify a suitable quantum state that not only displays the effect of de-cohering noise during its time evolution but can also robustly detect the presence of an external magnetic field. Many quantum states exhibit signatures of a magnetic field during their dynamics, but we focus on finding one that remains sensitive to external magnetic field despite the influence of significant decoherence. To achieve this, we consider a realistic exchange interaction and noise model, and identify a maximally coherent state that, even in the presence of strong noise, can effectively detect a magnetic field through its time evolution dynamics.  

\section{Exchange interaction and magnetic field}

We consider a Heisenberg interaction between the two spins in our device, where the exchange interaction occurs equally along the \(X\), \(Y\), and \(Z\) directions. The Hamiltonian for this exchange coupling is given by Eq.~\ref{E1}.

\begin{eqnarray}
   H_{EX} 
           &= J\left( \sigma_X \otimes \sigma_X+\sigma_Y \otimes \sigma_Y+\sigma_Z \otimes \sigma_Z \right), 
    \label{E1}
\end{eqnarray}
where \(J\) represents the coupling strength, and \(\sigma_i\) for \(i \in \{X, Y, Z\}\) are the Pauli matrices. The Heisenberg exchange interaction has been a fundamental mechanism for implementing two-qubit gates since the early days of quantum computing across multiple quantum platforms, including quantum dots, and donor-atom nuclear spins~\cite{PhysRevA.57.120,DiVincenzo2000}. Quantum computing is even possible with always-on Heisenberg exchange interaction~\cite{PhysRevLett.90.247901}. As the field continues to advance, Heisenberg exchange coupling remains a cornerstone of entangling qubits and performing essential quantum operations~\cite{Keith2022,PhysRevB.105.245413}.

\noindent We use a two-qubit spin system for magnetic field detection. We expose the device to a magnetic field along the \(Z\) direction. The effect of this magnetic field on the two spins can be described by the Hamiltonian in Eq.~\ref{E2}.

\begin{eqnarray}
H_M &= m\left( \sigma_Z \otimes I 
+ I \otimes \sigma_Z \right), 
\label{E2}
\end{eqnarray}
where, \(m\) is the magnetic field strength. The total Hamiltonian \(H\) of the two-qubit spin system in the presence of the magnetic field along the \(Z\) direction is therefore the sum of the exchange interaction Hamiltonian \(H_{EX}\) and the magnetic field Hamiltonian \(H_M\), as given by Eq.~\ref{E3}.

\begin{eqnarray}
H &= H_{EX}+H_M. 
\label{E3}
\end{eqnarray}

\noindent Our goal here, is to detect the strength of the magnetic field \(m\) by measuring the quantum coherence and return probability of the system after it has evolved under the influence of this total Hamiltonian \(H\). By analyzing how the coherence and return probability changes over time, we can infer the magnitude $m$ of the magnetic field.

\section{Noise model and initial state}\label{noise}

We consider a noise model where the noise impacts the coupling strength of the exchange Hamiltonian. Specifically, we model the coupling constant \(J\) as a random variable drawn from a Gaussian distribution with a mean of \(J_0\) and a standard deviation of \(\epsilon\).

\begin{eqnarray}
    J\in f(J|J_0,\epsilon)=\frac{1}{\epsilon \sqrt{2\pi}}e^{-\frac{1}{2}\left(\frac{J-J_0}{\epsilon}\right)^2}{.}
\end{eqnarray}

\noindent In the ideal case, the coupling strength \(J\) would be exactly equal to \(J_0\). However, due to the presence of noise, \(J\) can deviate from \(J_0\) and take on values in its vicinity, with the probability of these deviations following an exponential decay as described by the Gaussian distribution. The standard deviation \(\epsilon\) in the Gaussian distribution quantifies the effect of noise: a larger \(\epsilon\) implies a greater likelihood that \(J\) will significantly deviate from \(J_0\). Thus, \(\epsilon\) directly reflects the strength of the noise -- higher noise levels lead to a wider spread of possible \(J\) values around the mean \(J_0\).

\noindent The noise model considered here is commonly referred to as the quasi-static charge noise in the literature~\cite{Keith2022,PhysRevB.105.245413}. At high noise levels, charge noise is typically modeled by assuming a Gaussian distribution for the exchange interaction, an approach that remains valid to date~\cite{Keith2022,PhysRevB.105.245413}. Reducing charge noise is critical for achieving high-fidelity gate operations, and significant progress has been made in its analysis and mitigation~\cite{PhysRevA.100.022337,Yang2019,PhysRevB.108.045305,PaqueletWuetz2023,Elsayed2024}. {We employed Gaussian distribution to model the fluctuations in the parameter \( J \) during our analysis, as it is the most commonly used approach~\cite{Keith2022,PhysRevB.105.245413}. Other symmetric distributions, such as uniform and Cauchy-Lorentz distribution, produced qualitatively similar results for the analysis presented in this paper.
}

\noindent To study the dynamics of the two-spin system, we first need to initialize it in a specific state. Since our goal is to measure quantum coherence, it is beneficial to start with a state that is maximally coherent. A maximally coherent state~\cite{Bai2015MaximallyCS,PhysRevA.93.032326} is one that exhibits the greatest possible superposition, allowing us to observe the effects of decoherence most clearly. For this reason, we choose the initial state of the system to be \(\psi(0) = |+-\rangle\), which is a two-qubit state with equal superposition of the computational basis states.
This choice of initial state ensures that the system begins in a highly coherent state, making it ideal for analyzing how coherence evolves and degrades over time due to the dynamics governed by the Hamiltonian.

\section{Dynamics in absence of magnetic field}

In this section, we first explore the coherence dynamics of the two-spin system under the exchange Hamiltonian \( H_{EX} \) without considering the effect of the magnetic field. The impact of the magnetic field will be addressed in the subsequent section.

\noindent The time evolution of the system governed by the exchange Hamiltonian is described by the unitary time evolution operator \( U_{H_{EX}} \), which is given by:
$U_{H_{EX}} = e^{-iH_{EX}t}$.
Due to the specific form of the exchange Hamiltonian \( H_{EX} \) considered in our study, the time evolution operator can be explicitly written as:

\begin{equation}
U_{H_{EX}} = \left( \begin{array}{cccc} 
e^{-iJt} & 0 & 0 & 0 \\ 
0 & \frac{1}{2}\left(e^{-iJt}+e^{i3Jt}\right) & \frac{1}{2}\left(e^{-iJt}-e^{i3Jt}\right) & 0 \\ 
0 & \frac{1}{2}\left(e^{-iJt}-e^{i3Jt}\right) & \frac{1}{2}\left(e^{-iJt}+e^{i3Jt}\right) & 0 \\ 
0 & 0 & 0 & e^{-iJt} 
\end{array} \right){.}
\end{equation}

\noindent {The Heisenberg exchange interaction unitary \( U_{H_{EX}} \) is a powerful tool for implementing two-qubit gates in quantum computing. For instance, applying \( U_{H_{EX}} \) for a specific time \( t = \frac{\pi}{4J} \) results in the SWAP gate operation. This can be seen from the matrix representation of the unitary operator:
\begin{eqnarray}
 U_{H_{EX}}\left(\frac{\pi}{4J}\right) = e^{-iH_{EX}\frac{\pi}{4J}} &= \left( \begin{array}{cccc} 
e^{-i\frac{\pi}{4}} & 0 & 0 & 0 \\ 
0 & 0 & e^{-i\frac{\pi}{4}} & 0 \\ 
0 & e^{-i\frac{\pi}{4}} & 0 & 0 \\ 
0 & 0 & 0 & e^{-i\frac{\pi}{4}} 
\end{array} \right) 
\\ \nonumber
& = e^{-i\frac{\pi}{4}}\left( \begin{array}{cccc} 
1 & 0 & 0 & 0 \\ 
0 & 0 & 1 & 0 \\ 
0 & 1 & 0 & 0 \\ 
0 & 0 & 0 & 1  
\end{array} \right){,}
\\ \nonumber
\end{eqnarray}
where the second matrix is the SWAP gate. Thus, \( U_{H_{EX}}\left(\frac{\pi}{4J}\right) \) corresponds to the SWAP gate up to a global phase factor \( e^{-i\frac{\pi}{4}} \). This demonstrates the significance of the Heisenberg exchange interaction for implementing essential quantum gates like SWAP, making it a crucial interaction in quantum information processing.}

\noindent Next, we apply the time evolution operator \( U_{H_{EX}} \) to the initial density matrix \( \rho(0) \), which corresponds to the maximally coherent initial state. The density matrix after time evolution is given by:

\begin{equation}
\rho(t) = U_{H_{EX}} \rho(0) U_{H_{EX}}^{\dagger}{.}
\end{equation}

\noindent Substituting the explicit form of the initial state \( \rho(0) = |\psi(0)\rangle \langle \psi(0)| \), where \( |\psi(0)\rangle = |+-\rangle \), and applying the time evolution operator, we obtain the following evolved density matrix:

\begin{equation}
\rho(t) = \frac{1}{4} \left( \begin{array}{cccc} 
1 & -e^{-i4Jt} & e^{-i4Jt} & -1 \\ 
-e^{i4Jt} & 1 & -1 & e^{i4Jt} \\ 
e^{i4Jt} & -1 & 1 & -e^{i4Jt} \\ 
-1 & e^{-i4Jt} & -e^{-i4Jt} & 1 
\end{array} \right){.}
\end{equation}

\noindent This resulting density matrix describes the system's state after evolving under the exchange Hamiltonian \( H_{EX} \) for a time \( t \), in the absence of any noise. It represents the "ideal" case where no errors are present. However, in practical scenarios, the coupling strength \( J \) is not fixed but subject to fluctuations due to noise, as discussed in section~\ref{noise}. To account for this, we consider the noise-affected coupling strength \( J \) as a random variable distributed according to a Gaussian distribution with mean \( J_0 \) and standard deviation \( \epsilon \). To understand the impact of this noise, we compute the average time-evolved density matrix \( \overline{\rho(t)} \) by averaging over all possible values of \( J \), weighted by the probability density function \( f(J|J_0,\epsilon) \). The average density matrix is given by:

\begin{equation}
\overline{\rho(t)} = \int f(J|J_0,\epsilon) \,\rho(t) 
\, dJ {.}
\end{equation}
{Without loss of generality and for simplicity in calculations, we set \( J_0 = 1 \).} After performing this integration for all the entries of the density matrix ${\rho(t)}$, the average time-evolved density matrix, which incorporates the effect of the noise, is found to be:

\begin{equation}
\overline{\rho(t)} = \frac{1}{4} \left( \begin{array}{cccc} 
1 & -e^{-i4t - 8 \epsilon^2 t^2} & e^{-i4t - 8 \epsilon^2 t^2} & -1 \\ 
-e^{i4t - 8 \epsilon^2 t^2} & 1 & -1 & e^{i4t - 8 \epsilon^2 t^2} \\ 
e^{i4t - 8 \epsilon^2 t^2} & -1 & 1 & -e^{i4t - 8 \epsilon^2 t^2} \\ 
-1 & e^{-i4t - 8 \epsilon^2 t^2} & -e^{-i4t - 8 \epsilon^2 t^2} & 1 
\end{array} \right){.}
\end{equation}

\noindent We then calculate the $l_1$-norm: 
\begin{eqnarray}
    \label{eq12}C_{l_1}\left(\overline{\rho(t)}\right)={\color{black}\frac{1}{4-1}}\sum_{i\ne j}|\overline{\rho_{ij}(t)}|=\frac{1}{3}\left(1+2e^{-8\epsilon^2 t^2}\right).
\end{eqnarray}
This expression shows that as the noise parameter \( \epsilon \) increases, the coherence decays more rapidly, highlighting the effect of noise on quantum systems. We also calculate the return probability of the time evolved state to the initial state: 
\begin{eqnarray}
P_R\left(\overline{\rho(t)}\right)=\langle \psi(0)|\overline{\rho(t)}|\psi(0)\rangle=\frac{1}{2}\left(1+cos(4t)e^{-8\epsilon^2 t^2}\right). \label{RP}
\end{eqnarray} 
The return probability shows a damped oscillatory behavior.

\section{Dynamics in presence of magnetic field}

In this section, we consider the impact of the magnetic field i.e., we consider the total Hamiltonian $H$. The time evolution of the system governed by the total Hamiltonian is described by the unitary time evolution operator \( U_{H} \), which is given by: $U_{H} = e^{-iHt}$.
The time evolution operator can be explicitly written as:

\begin{equation}
U_{H} = \left( \begin{array}{cccc} 
e^{-i(J+2m)t} & 0 & 0 & 0 \\ 
0 & \frac{1}{2}\left(e^{-iJt}+e^{i3Jt}\right) & \frac{1}{2}\left(e^{-iJt}-e^{i3Jt}\right) & 0 \\ 
0 & \frac{1}{2}\left(e^{-iJt}-e^{i3Jt}\right) & \frac{1}{2}\left(e^{-iJt}+e^{i3Jt}\right) & 0 \\ 
0 & 0 & 0 & e^{-i(J-2m)t} 
\end{array} \right){.}
\end{equation}
Next, we apply the time evolution operator \( U_{H} \) to the initial density matrix \( \rho(0) \), which corresponds to the maximally coherent initial state. The density matrix after time evolution is given by:

\begin{equation}
\rho_m(t) = U_{H} \rho(0) U_{H}^{\dagger}{.}
\end{equation}
Applying the time evolution operator, we obtain the evolved density matrix:

\begin{equation}
\rho_m(t) = \frac{1}{4} \left( \begin{array}{cccc} 
1 & -e^{-i(4J+2m)t} & e^{-i(4J+2m)t} & -e^{-i4mt} \\ 
-e^{i(4J+2m)t} & 1 & -1 & e^{i(4J-2m)t} \\ 
e^{i(4J+2m)t} & -1 & 1 & -e^{i(4J-2m)t} \\ 
-e^{i4mt} & e^{-i(4J-2m)t} & -e^{-i(4J-2m)t} & 1 
\end{array} \right){.}
\end{equation}
This density matrix describes the system's state after evolving under the total Hamiltonian \( H \) for a time \( t \), in the absence of any noise. Considering the impact of the noise, we compute the average time-evolved density matrix \( \overline{\rho(t)} \) by averaging over all possible values of \( J \), weighted by the probability density function \( f(J|J_0,\epsilon) \). The average density matrix is given by:

\begin{equation}
\overline{\rho_m(t)} = \int f(J|J_0,\epsilon) \, \rho_m(t) \, dJ{.}
\end{equation}
After performing this averaging, the average time-evolved density matrix, which incorporates the effect of the noise, is found to be:

{
\begin{equation}\label{Eq20}
\overline{\rho_m(t)} = \frac{1}{4} \left( \begin{array}{cccc} 
1 & -e^{-iat - c} & e^{-iat - c} & -e^{-i4mt} \\ 
-e^{iat - c} & 1 & -1 & e^{ibt - c} \\ 
e^{iat - c} & -1 & 1 & -e^{ibt - c} \\ 
-e^{i4mt} & e^{-ibt - c} & -e^{-ibt - c} & 1 
\end{array} \right),
\end{equation}
where $a=(4+2m)$, $b=(4-2m)$, and} $c = 8 \epsilon^2 t^2$.
We then calculate the \( l_1 \)-norm of the average time evolved density matrix: 
\begin{eqnarray}
    \label{eq19}C_{l_1}\left(\overline{\rho_m(t)}\right)=\frac{1}{3}\left(1+2e^{-8\epsilon^2 t^2}\right){.}
\end{eqnarray}
This expression shows that there is no effect of magnetic field in $l_1$-norm. We also calculate the return probability of the time evolved state to the initial state: 
\begin{eqnarray}
P_R\left(\overline{\rho_m(t)}\right)=\frac{3}{8}+ \frac{1}{8}\cos(4mt)+\frac{1}{2}\cos(2mt)\cos(4t)e^{-8\epsilon^2 t^2} {.}\label{RP_m}
\end{eqnarray}
In the above expression, there is an additional non-damping oscillatory part in the return probability in presence of magnetic field.

\begin{figure}
    \centering
    \includegraphics[width=0.75\linewidth]{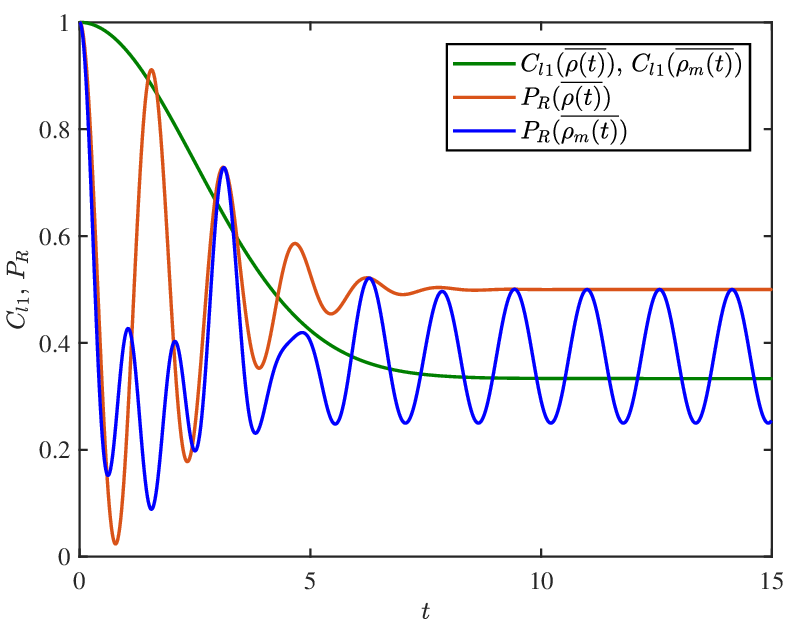}
    \caption{{Dynamics of the \( l_1 \)-norm and return probability with and without an external magnetic field. The green line represents the \( l_1 \)-norm, which remains unchanged regardless of the magnetic field. The orange and {blue} lines correspond to the return probabilities in the absence and presence of the magnetic field, respectively. Here in this figure, we considered the magnetic field strength $m=1$ and the strength of quasi-static charge noise $\epsilon=0.1$.}}
    \label{compare}
\end{figure}

\noindent By comparing Eq.~\ref{RP_m} with Eq.~\ref{RP}, we observe a clear difference in the behavior of the return probability with and without the magnetic field. At higher times $t$, when a magnetic field is present, the return probability, denoted by $P_R(\overline{\rho_m(t)})$, exhibits oscillatory behavior, while in the absence of a magnetic field, the return probability, denoted as $P_R(\overline{\rho(t)})$, remains almost stationary. Specifically, with the magnetic field, we see a non-damping oscillatory term $\frac{1}{8}\cos(4mt)$ in $P_R(\overline{\rho_m(t)})$, where the oscillations continue without fading over time. In contrast, the oscillations of $P_R(\overline{\rho(t)})$ in the absence of the magnetic field decay exponentially. Noise causes decoherence but does not affect the \({{\color{black}\frac{1}{8}\cos(4mt)}}\) term of the magnetic-field-induced oscillations, allowing for robust magnetic field detection. However, note that the \({{\color{black}\cos(2mt)}}\) component of the magnetic-field-induced oscillations is damped by decoherence. 
Additionally, the frequency of the non-damping oscillations is $4m$, which is directly proportional to the magnetic field strength $m$. This relationship allows us to determine the magnetic field strength by measuring the frequency of the non-damping oscillations in the return probability. {We see in Fig.~\ref{compare} that for sufficiently large times (\( t > 10 \)), the term \( \frac{1}{2}\cos(2mt)\cos(4t)e^{-8\epsilon^2 t^2} \) of Eq.~\ref{RP_m} becomes negligible, effectively approximating to zero. As a result, the period of oscillation \( T \) is primarily determined by the term \( \cos(4mt) \), yielding \( T \approx \frac{2\pi}{4m} \). For a field strength \( m = 1 \), this simplifies to \( T \approx \frac{\pi}{2} \).} On the other hand, it is not possible to use the $l_1$-norm to measure the magnetic field strength, as both $l_1(\overline{\rho(t)})$ and $l_1(\overline{\rho_m(t)})$ are identical, regardless of the presence or absence of the magnetic field.\\
{Coherence fundamentally refers to the preservation of quantum superposition along the computational or measurement basis. In our work, decoherence signifies the breakdown of the quantum superposition of the initial state in the \(\sigma_z\) basis. Due to the influence of the assumed quasi-static charge noise, the coherence of the maximally coherent initial state decreases but does not fully vanish during the dynamics.} Even at $t=\infty$, the state retains a non-zero coherence according to Eq.~\ref{eq12} and Eq.~\ref{eq19}. The residual coherence indicates that while the noise disrupts the purity and superposition of the initial state along the computational basis, transforming it into a mixed state, {it} does not drive the state to a maximally decohered mixed state. Fig.~\ref{compare} illustrates the dynamics of the \( l_1 \)-norm and return probability, both in the presence and absence of an external magnetic field, governed by the noisy Heisenberg exchange interaction Hamiltonian.

\noindent{Measuring the return probability to the initial state \( |+-\rangle \) requires performing measurements in the computational \(\sigma_z\) basis after applying a Hadamard gate to both qubits or, equivalently, measuring directly in the \(\sigma_x\) basis. Pauli measurements are the most commonly implemented measurements in quantum computing devices. In contrast, measuring the \( l_1 \)-norm is more challenging, as it necessitates full quantum state tomography.}\\
\noindent The choice of the initial state $\psi(0)=|+-\rangle$ is critical for studying the impact of the de-cohering noise in our model. This state is maximally coherent in the computational basis and exhibits the effects of quasi-static charge noise during its time evolution under the Heisenberg exchange coupling Hamiltonian. The presence of an external magnetic field induces oscillations in the evolution of this initial state. Over time, the quasi-static charge noise leads to decoherence, transforming the pure initial state into a mixed state. Despite this, a portion of the magnetic field-induced oscillations persists in the de-cohered mixed state, enabling magnetic field detection even in the presence of significant noise. It is important to note that some initial states, such as $|++\rangle$, remain unaffected by decoherence during their dynamics. However, in this study, our focus is on an initial state that both displays the effects of decoherence and retains sensitivity to magnetic field detection.

\section{Limitations of the method in the presence of depolarizing noise}

{Our proposed method for magnetic field detection performs effectively under the influence of quasi-static charge noise, which is the most prominent noise during two-qubit gate operations mediated by the exchange interaction. However, our analysis assumes the complete absence of other types of noise, which is not a realistic assumption in practical scenarios. The presence of additional noise sources could potentially disrupt the accuracy of our magnetic field detection method. To address this, we additionally introduce the depolarizing noise model into the system as a representative of other possible noises.}


\noindent{Depolarizing noise is often used as a simplified model because it captures the isotropic nature of many realistic noise sources in spin qubit systems. While real-world noise may have some anisotropy, depolarizing noise provides a tractable and general framework for studying the effects of decoherence on quantum systems~\cite{nielsen_chuang_2010,PhysRevA.102.052608,PhysRevA.105.052608,PhysRevX.13.041013,hu2024scalabilityenhancementquantumcomputing,PRXQuantum.4.040341,PhysRevX.13.041022}.
The depolarizing channel describes a process where a quantum state (represented by a density matrix \( \rho \)) is replaced by a completely mixed state with probability \( 1 - p \), and remains unchanged with probability \( p \). Here, \( p \) is called the purity preserving factor or the probability of no error.
For a two-qubit system, the depolarizing channel \( \mathcal{D}(\rho) \) acting on a density matrix \( \rho \) is defined as:
\begin{equation}\label{Eq_depolarizing}
\mathcal{D}(\rho) = p \, \rho + (1 - p) \frac{I}{4},
\end{equation}
where:
\( \rho \) is the input density matrix of the qubit,
\( I \) is the \( 4 \times 4 \) identity matrix (representing the completely mixed state in the 4-dimensional Hilbert space),
\( p \) is the purity preserving factor (\( 0 \leq p \leq 1 \)).
We assume that the impact of depolarizing noise on the qubits intensifies with the duration of the evolution under the Heisenberg exchange interaction, as supported by previous studies~\cite{PhysRevA.105.012432,PhysRevResearch.5.033055}. This effect is quantified by the purity preservation factor \( p \), which decays exponentially over time according to:
\begin{equation}\label{preserving}
p = \exp(-\gamma t),
\end{equation}
where \( \gamma \) is the decay rate of the purity preservation factor. A higher value of \( \gamma \) corresponds to a stronger influence of the depolarizing error channel, leading to faster degradation of the qubit's purity.}
{With the inclusion of depolarizing noise, the initial state, which evolves under the Heisenberg exchange interaction and is subjected to quasi-static charge noise, further experiences the effects of a depolarizing channel.} {As given in Eq.~\ref{Eq20}, the density matrix \( \overline{\rho_m(t)} \) represents the state after undergoing noisy Heisenberg interaction in the presence of an external magnetic field for an evolution time \( t \). The noisy Heisenberg interaction is modeled as a unitary evolution \( U_H \), followed by an averaging over the interaction strength \( J \), which is assumed to follow a Gaussian distribution. This evolved state, \( \overline{\rho_m(t)} \), then undergoes the depolarizing noise channel described in Eq.~\ref{Eq_depolarizing}, with the time-dependent purity-preserving factor defined in Eq.~\ref{preserving}. The resulting density matrix after passing through the depolarizing channel is given by Eq.~\ref{EqDp}.}

{
\begin{equation}\label{EqDp}
\mathcal{D}\left(\overline{\rho_m(t)}\right) = \frac{1}{4} \left( \begin{array}{cccc} 
1 & -pe^{-iat - c} & pe^{-iat - c} & -pe^{-i4mt} \\ 
-pe^{iat - c} & 1 & -p & pe^{ibt - c} \\ 
pe^{iat - c} & -p & 1 & -pe^{ibt - c} \\ 
-pe^{i4mt} & pe^{-ibt - c} & -pe^{-ibt - c} & 1 
\end{array} \right),
\end{equation}}
\noindent where $a=(4+2m)$, $b=(4-2m)$, and $c = 8 \epsilon^2 t^2$. {Notably, after passing through the depolarizing channel, all off-diagonal elements of the density matrix are scaled by a factor \( p \). Consequently, the \( l_1 \)-norm of quantum coherence is also scaled by \( p \), leading to the expression:
\begin{eqnarray}\label{eq19D}
C_{l_1}\left(\mathcal{D}\left(\overline{\rho_m(t)}\right)\right) = p \frac{1}{3} \left(1 + 2 e^{-8\epsilon^2 t^2}\right) = e^{-\gamma t} \frac{1}{3} \left(1 + 2 e^{-8\epsilon^2 t^2}\right).
\end{eqnarray}}
\noindent{From Eq.~\ref{eq19D}, we observe that as \( t \to \infty \), the \( l_1 \)-norm of coherence tends to zero, indicating that the state undergoes maximal decoherence due to the depolarizing noise.} {The return probability of the system to its initial state is given by:
\begin{eqnarray}
P_R\left(\mathcal{D}\left(\overline{\rho_m(t)}\right)\right) = \frac{1}{4} + e^{-\gamma t} \left( \frac{1}{8} + \frac{1}{8} \cos(4mt) + \frac{1}{2} \cos(2mt) \cos(4t) e^{-8\epsilon^2 t^2} \right). \label{Rp_mD}
\end{eqnarray}}
{From Eq.~\ref{Rp_mD}, we see that the presence of depolarizing noise leads to the exponential damping of the previously non-decaying oscillatory term \( \frac{1}{8} \cos(4mt) \), as it is now multiplied by \( e^{-\gamma t} \). This indicates that the return probability oscillations are suppressed over time due to the depolarizing noise.}

\begin{figure}
    \centering
    \includegraphics[width=0.75\linewidth]{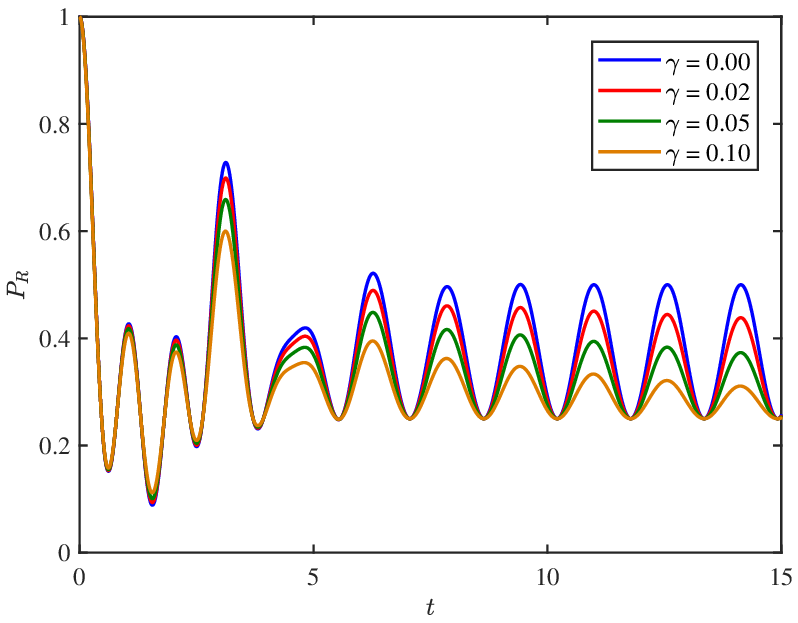}
    \caption{{Return probability dynamics in the presence of an external magnetic field with strength \( m = 1 \), quasi-static charge noise with strength \( \epsilon = 0.1 \), and depolarizing noise with different decay rates \( \gamma \). The figure demonstrates that while depolarizing noise dampens oscillations of return probability, the oscillation frequency remains unaffected.}}
    \label{fig:ReturnProbability_bothnoise}
\end{figure}

\noindent{Fig.~\ref{fig:ReturnProbability_bothnoise} illustrates the return probability dynamics for different values of \( \gamma \). We observe that depolarizing noise suppresses oscillation amplitude. For small values of \( \gamma \), this effect is minor, but as \( \gamma \) increases, the oscillations become significantly damped, reducing the visibility of oscillations in the return probability. Importantly, despite the presence of depolarizing noise, the oscillation frequency remains unchanged, suggesting that our magnetic field detection method remains robust under low-strength depolarizing noise.}

\noindent{To quantitatively understand how realistic noise, such as depolarizing noise (in addition to quasi-static charge noise), affects the precision of our magnetic field detection method, we compute the Quantum Fisher Information (QFI) of the density matrix \( \mathcal{D}(\overline{\rho_m(t)}) \), (given in Eq.~\ref{EqDp}), with respect to the detection parameter \( m \), which represents the external magnetic field.}


\noindent{In classical statistics, Fisher information quantifies how much information a random variable carries about an unknown parameter. {It plays a crucial role in estimation theory, providing a lower bound on the variance of an estimator, known as the \textit{Cramér-Rao bound}~\cite{Cramer_1946,Rao_1945}.
Quantum Fisher Information extends this concept to quantum systems, where measurements are inherently probabilistic and depend on observables~\cite{doi:10.1142/S0219749909004839,doi:10.1142/9789814338745_0015}.} In quantum parameter estimation, a parameter \( m \) is encoded into a quantum state \( \rho(m) \) through a parameter-dependent evolution. The precision of estimating \( m \) is constrained by the \textit{Quantum Cramér-Rao Bound (QCRB)}, which directly depends on the QFI. A higher QFI value implies a lower variance in the estimator, leading to greater precision in the parameter estimation~\cite{PhysRevA.95.052320,Liu_2020}}.
\noindent{The QCRB establishes a fundamental limit on the precision of parameter estimation, given by:
\begin{equation}
    \sigma_m \geq \frac{1}{\sqrt{\nu \times QFI}},
\end{equation}
where:
\begin{itemize}
    \item \( \sigma_m \) is the standard deviation (uncertainty) of the estimator for the parameter \( m \). A lower {\(\sigma_m\)} indicates higher estimation precision.
    \item \( \nu \) is the number of repeated measurements. Increasing \( \nu \) improves precision by reducing statistical fluctuations.
    \item \( QFI \) is the Quantum Fisher Information, which quantifies the sensitivity of the quantum state to changes in \( m \). A larger \( QFI \) means that small variations in \( m \) induce more significant changes in the quantum state, enhancing the estimator’s sensitivity.
\end{itemize}}

\noindent{By analyzing the QFI under different noise conditions, we can assess the fundamental limits on the precision of our magnetic field detection method and determine how noise sources impact the effectiveness of the estimation process.}

\noindent{For a density matrix \( \rho \), the QFI with respect to a single parameter \( m \) is given by:
\begin{equation}
{QFI}{(\rho)_{m}} = \sum_{i,j=0}^{d-1} \frac{2 |\langle \lambda_i | \partial_m \rho | \lambda_j \rangle|^2}{\lambda_i + \lambda_j}, \quad {with} \quad \lambda_i + \lambda_j \neq 0,
\label{formula: QFI}    
\end{equation}
where  \(d\), \( |\lambda_i\rangle \) and \( \lambda_i \) are the dimensionality, eigenvectors and eigenvalues of \( \rho \), respectively, and \( \partial_m \rho \) denotes the derivative of \( \rho \) with respect to the parameter \( m \)~\cite{Liu_2020}.}

\noindent{Using Eq.~\ref{formula: QFI}, we numerically compute the QFI dynamics of the density matrix \( \mathcal{D}\left(\overline{\rho_m(t)}\right) \) as a function of the magnetic field strength \( m \) for different values of the depolarizing decay rate \( \gamma \).}
\begin{figure}
\centering\includegraphics[width=0.75\linewidth]{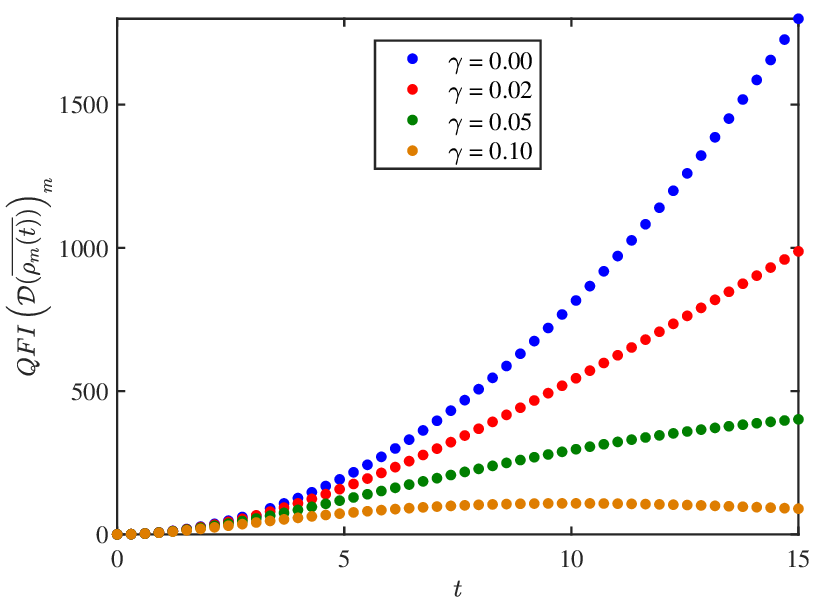}
    \caption{{QFI dynamics under depolarizing noise. The case \( \gamma = 0 \) corresponds to an ideal scenario without depolarizing noise, where the QFI continuously increases, enhancing sensor performance. However, as \( \gamma \) increases, the QFI exhibits an initial rise followed by a decline, indicating a reduction in estimation precision. Since QFI is inversely related to the variance of the estimated parameter, this result highlights the detrimental impact of depolarizing noise on sensor efficiency.}}
    \label{fig:QFI}
\end{figure}
{Fig.~\ref{fig:QFI} illustrates the effect of depolarizing noise on QFI. In the absence of depolarizing noise (\( \gamma = 0 \)), the QFI increases monotonically with time \( t \), indicating that the precision of magnetic field estimation improves for longer evolution times. However, for nonzero values of \( \gamma \), the QFI initially increases but eventually decreases. A higher depolarizing decay rate leads to a lower QFI and a faster decline, demonstrating that strong depolarizing noise degrades the precision of our magnetic field detection.}

\noindent{Note that the two-qubit system we considered is a spin-based quantum computing device that operates at sufficiently low temperatures, adequately shielded from environmental disturbances~\cite{PhysRevA.57.120,Jeong2024}. We investigate whether this device can be utilized to sense static magnetic fields of the order of the exchange interaction strength. The system would not function {as} effectively in the presence of significant decoherence.}

\section{Comparative Discussion with Existing Methods}

{So far, we have discussed the proposed magnetic field detection method in detail, including its underlying principles and limitations. {In this work, we study a two-qubit spin system with Heisenberg exchange interaction, without specifying the substrate. As discussed in the limitations, very low environmental noise is a necessity for the method to work properly.} In this section, we provide a comparative analysis of this method with other established techniques from the literature.}

\noindent{Spins are natural candidates for quantum sensing of magnetic fields due to their inherent two-level structure, long coherence times, and sensitivity to magnetic fields~\cite{ROSE201732,PhysRevApplied.20.044045,Patrickson2024}. Examples of such spin-based sensors include neutral atom sensors, solid-state spin sensors, and nitrogen-vacancy (NV) centers in diamond \cite{RevModPhys.89.035002,https://doi.org/10.1002/mrm.21624,Kitching2011}. Among these, NV centers in diamond have emerged as a particularly promising platform for quantum sensing due to their exceptional coherence properties and sensitivity \cite{annurev:/content/journals/10.1146/annurev-physchem-040513-103659,Rondin_2014,Huxter2022}.}

\noindent{One of the widely used quantum sensing scheme is Ramsey interferometry \cite{RevModPhys.89.035002,RevModPhys.90.035005}, which relies on two key state transformations to control the coherent evolution of a probe qubit under an external influence. Initially, the qubit is prepared in a coherent superposition state:
\[
\psi(0) = \frac{1}{\sqrt{2}} (|0\rangle + |1\rangle).
\]
This state evolves over a time \( t \) under the influence of an external signal, acquiring a phase \( \theta(t) \):
\[
\psi(t) = \frac{1}{\sqrt{2}} (|0\rangle + \exp(-i\theta(t))|1\rangle).
\]
The accumulated phase is given by:
\[
\theta(t) = \int_0^t [\omega_0 + \delta\omega(t')] \, dt',
\]
where \( \hbar\omega_0 \) is the static energy gap between the qubit's energy levels, and \( \hbar\delta\omega(t') \) represents a small, time-dependent modulation induced by the external signal. After this evolution, a second transformation reorients the qubit back to its original basis, followed by a projective measurement onto the initial state. The transition probability \( p \) is measured as a function of time, and the signal is extracted from the probability difference \( \delta p = p - p_0 \), where \( p_0 \) is the probability in the absence of the signal.}

\noindent{The proposed method shares a similarity with Ramsey interferometry in that both rely on measuring the return probability of a quantum state. However, the mechanisms differ significantly. In Ramsey interferometry, the return probability changes due to the phase accumulation during evolution in the presence of an external field. In the proposed method, the return probability is influenced by the evolution under a noisy Heisenberg exchange interaction, and its oscillations are modulated by the presence of an external magnetic field.} 

\noindent{There are several key differences between the proposed method and Ramsey interferometry:
\begin{itemize}
    \item \textbf{Interaction}: The proposed method incorporates the Heisenberg exchange interaction between the two qubits, which is a widely used interaction for implementing two-qubit gates.
    \item \textbf{Initial State}: We initialize the system in a specific maximally coherent state, which simplifies the computation of return probabilities using only Pauli measurements.
    \item \textbf{Noise}: The proposed method explicitly accounts for noise in the exchange interaction, making it more robust to imperfections in real-world implementations.
\end{itemize}}

\noindent{The central question addressed in this work is whether an imperfect noisy two-qubit gate implementation can be used to detect a static external magnetic field with a strength comparable to the two-qubit exchange interaction. The results demonstrate that such a detection is indeed feasible, offering a novel approach to magnetic field sensing that leverages the inherent properties of noisy quantum systems.}

\section{Conclusion}

In this work, we have proposed a method for detecting magnetic fields using a two-qubit spin  system governed by noisy Heisenberg exchange interaction. We demonstrated that the return probability of the system exhibits characteristic oscillations in response to an external magnetic field, with the oscillation frequency directly proportional to the field strength. By incorporating a noise model that accounts for quasi-static charge noise~\cite{Keith2022,PhysRevB.105.245413}, we showed that while the noise leads to decoherence, it does not affect a part of the magnetic-field-induced oscillations in the return probability. This clear distinction between the effects of noise and the magnetic field highlights the robustness of the system for magnetic field detection. {However, considering only quasi-static charge noise in the system is not realistic. We also discussed the limitations of the proposed method in the presence of depolarizing noise, which serves as a representative model for other possible noise sources.}

\noindent Our study offers a practical approach to magnetic field detection by utilizing Heisenberg exchange-coupled qubits, which are adaptable across various quantum computing platforms~\cite{PhysRevA.57.120,DiVincenzo2000}. The results demonstrate that magnetic field detection is feasible even in noisy environments, underscoring the potential of noisy two-qubit systems for quantum sensing applications. This work contributes to the expanding field of quantum technologies, presenting a noise-resilient method for precision sensing in real-world conditions.

\section*{Acknowledgments}
GB is supported by the National Science and Technology Council, Taiwan, under Grant No. NSTC 
112-2112-M-032-008-MY3, 
112-2811-M-032-005-MY2. 

\section*{References}
\bibliographystyle{iopart-num}
\bibliography{PhD_references}

\end{document}